# Explainable machine learning identifies multi-omics signatures of muscle response to spaceflight in mice

Authors

Kevin Li[1,2*], Riya Desai[3*], Ryan T. Scott[1,9], Joel Ricky Steele[4,8,9], Meera Machado[5], Samuel Demharter[5], Adrienne Hoarfrost[6], Jessica L. Braun[7], Val A. Fajardo[7], Lauren M. Sanders[8,9], Sylvain V. Costes[9]

1. KBR, Moffett Field, CA, USA
2. NASA Space Life Sciences Training Program, Moffett Field, CA, USA
3. College of Letters and Science, University of California at Davis, Davis, CA, USA
4. Monash Proteomics and Metabolomics Platform, Monash Biomedicine Discovery Institute, Monash University, Clayton, Victoria 3800, Australia
5. Abzu ApS, Denmark
6. Department of Marine Sciences, University of Georgia, Athens, GA, USA
7. Department of Kinesiology, Centre for Bone and Muscle Health, Brock University, Canada
8. Blue Marble Space, Seattle, WA, USA
9. Space Biosciences Division, NASA Ames Research Center, Moffett Field, CA, USA

* Co first author

Abstract

The adverse effects of microgravity exposure on mammalian physiology during spaceflight necessitate a deep understanding of the underlying mechanisms to develop effective countermeasures. One such concern is muscle atrophy, which is partly attributed to the dysregulation of calcium levels due to abnormalities in SERCA pump functioning. To identify potential biomarkers for this condition, multi-omics data and physiological data available on the NASA Open Science Data Repository (osdr.nasa.gov) were used, and machine learning methods were employed. Specifically, we used multi-omics (transcriptomic, proteomic, and DNA methylation) data and calcium reuptake data collected from C57BL/6J mouse soleus and tibialis anterior tissues during several 30+ day-long missions on the international space station. The QLattice symbolic regression algorithm was introduced to generate highly explainable models that predict either experimental conditions or calcium reuptake levels based on multi-omics features. The list of candidate models established by QLattice was used to identify key features contributing to the predictive capability of these models, with Acyp1 and Rps7 proteins found to be the most predictive biomarkers related to the resilience of the tibialis anterior muscle in space. These findings could serve as targets for future interventions aiming to reduce the extent of muscle atrophy during space travel.

## Introduction

Muscle atrophy, caused by prolonged exposure to microgravity conditions, is a major challenge faced by astronauts during spaceflight[1,2]. Although intense physical exercise is currently the main countermeasure, it requires a significant amount of time from each astronaut (2.5 hours per day, including equipment setup and breakdown), and even with exercise, the continuous exposure to microgravity cannot be fully offset.

It has been proposed that muscle atrophy may be at least partly explained by the dysregulation of cytoplasmic $Ca^{2+}$ levels due to abnormalities in the Sarco Endoplasmic Reticulum Calcium ATPase (SERCA) pump's ability to reuptake cytoplasmic $Ca^{2+}$ during muscle relaxation[1]. Mammalian muscles are broadly classified into two types: slow-twitch muscles composed predominantly of oxidative muscle fibers (e.g. postural muscles like the soleus [SOL], which is found in the calf and is important for resisting the pull of gravity) and fast-twitch muscles composed predominantly of glycolytic muscle fibers (e.g. explosive muscles like the tibialis anterior [TA], located in the shin). The SOL and TA are two of the primary muscles impacted by spaceflight, and previous studies have shown that both the murine SOL and TA will atrophy in response to microgravity exposure[1,3].

With respect to $Ca^{2+}$ handling, recent work has shown that during spaceflight $Ca^{2+}$ uptake is impaired in the SOL muscle, while being enhanced in the TA muscle, indicating that SERCA function is affected differently in the two muscle types[1]. However, the molecular mechanisms driving these aberrations in $Ca^{2+}$ reuptake by SERCA are not very well elucidated. Additionally, as $Ca^{2+}$ handling at the level of SERCA was not impaired in the TA, it is possible that there may be other molecular drivers of the muscle atrophy phenotype. Understanding these mechanisms at the molecular level is important for prevention and mitigation.

Machine learning methods are particularly effective in identifying patterns in complex biological data, particularly for discovering biomarkers in heterogeneous, high-dimensional, multi-omics datasets[4]. Compared to traditional statistical methods, machine learning (ML) methods are also less prone to distribution-specific effects[5], making them a promising alternative to classic systems biology. This is particularly important for space biological research, where small datasets are often combined to increase statistical power.

In this study, we present an ML-based approach to create a mapping between changes in multi-omics data (transcriptomic, proteomic, and epigenomic) and calcium reuptake in the SOL and TA muscles of mice that have been flown in space, compared to ground controls. To our knowledge, this approach has not been applied to this scientific question before.

When choosing a machine learning method for the purposes of gaining insight in biomedical research, it is important to consider two criteria: explainability/interpretability and generalizability. Many conventional state-of-the-art algorithms, such as neural networks, are seen as "black boxes" due to the low interpretability of the values and interactions of intermediate neurons deep within the network. Such algorithms thus have low explainability and are not ideal for research, where the ultimate goal is not performance, but rather acquiring a more sophisticated understanding of relationships between variables. Furthermore, a major challenge posed by multi-omics data in particular is the lack of generalizability of learned models due to the heterogeneity, high-dimensionality and low-sample-size (HDLSS) nature of the data. Highly expressive algorithms like neural networks will generate severely overfit models when trained on HDLSS data[6].

With these criteria in mind, we chose to use a recently-developed implementation of symbolic regression called QLattice, developed by Abzu ApS[7,8]. Symbolic regression attempts to find the true, concise mathematical function directly underlying the features' relationship to the target, which is much more interpretable than neural network architectures and less likely to be overfit. It does this by representing mathematical expressions as computational graphs, where the nodes represent variables or functions, and by exploring the possible architectures for these computational graphs. QLattice explores this graph space efficiently to find concise, interpretable, and accurate models, which make it a suitable tool for biomarker discovery[9].

We aimed to identify novel molecular drivers of spaceflight effects on muscle physiology using spaceflight mouse muscle data from the NASA Open Science Data Repository (OSDR) with omics data found in GeneLab[10,11] and $Ca^{2+}$ reuptake data found in Ames Life Sciences Data Archive (ALSDA)[12]. First, we trained QLattice to predict calcium reuptake levels of spaceflight and ground control mice using multi-omics features (genes, proteins, etc.), a regression task. Second, we trained QLattice to predict whether samples were from spaceflight or ground control mice, a classification task. We then identified features that contributed most to the predictive capabilities of the models output by QLattice; these features (i.e. genes, proteins, or epigenetic markers) are potential biomarkers that may provide mechanistic insight behind the spaceflight-induced muscle physiology effects, and serve as targets for future interventions aiming to reduce the extent of muscle atrophy during space travel.

## Materials and methods

### Data availability

The datasets used in this study were collected from the NASA Rodent Research 1 (RR-1) and 9 (RR-9) missions and are publicly available on the NASA OSDR (osdr.nasa.gov). RR-1 samples were from female C57BL/6J mice flown at 16-weeks of age for 37 days. RR-9 samples were from male C57BL/6J mice flown at 10-weeks of age for 35 days. Specifically, we used datasets OSD-104 (RR-1 multi-omics mouse SOL data)[13], OSD-105 (RR-1 multi-omics mouse TA data)[14], and OSD-488 (RR-1 and RR-9 calcium reuptake data)[15]. OSD-104 dataset consists of bulk RNA-sequencing (RNA-seq) and bisulfite sequencing DNA methylation data for SOL muscle samples collected from 6 space flown mice (FLT) and 6 ground control mice (GC) during the RR-1 mission[14]. OSD-105 dataset consists of bulk RNA-seq, bisulfite sequencing DNA methylation, and mass-spectrometry-based proteomics data for TA muscle samples, also collected from 6 FLT and 6 GC mice during RR-1[13]. OSD-488 dataset[1,15] originates from a study consisting of calcium reuptake data from female SOL muscle samples collected from 4 FLT and 4 GC during RR-1 mission; 10 FLT and 10 GC of SOL and TA male muscle samples during the RR-9 mission.

### Code availability

Due to the NASA Software Release requirements, the code for this study is not publicly Available.

### RNA sequencing data (RR-1 - SOL and TA muscles - female)

For RNA-seq RR-1 data, we started with the raw counts files available on OSDR from the GeneLab RNA-seq processing pipeline[16]. We filtered out lowly expressed genes with unreliable reads (i.e. we only kept genes that had at least 10 non-zero reads in at least 3 samples), resulting in a reduction of dimensionality from 55,536 genes to 15,848 genes for OSD-104 and 16,660 genes for OSD-105. We then applied a variance-stabilizing transformation (VST) using DESeq2 v1.34.0[17], which corrected for library size/sequencing depth and mitigated heteroskedasticity and skewed distributions. The mean-variance relationship and count distribution were checked post-normalization to confirm the effectiveness of preprocessing (Supplementary Figure 1a-d, Supplementary Figure 3a-d).

*Proteomics data (RR-1 - TA muscle only - female)*

The proteomics data for OSD-105 was collected in two runs of TMT-labeled mass spectrometry with a bridge channel in each run consisting of a pooled sample of all FLT and GC samples[13]. To ensure values were comparable when combining the runs, we used the ratio of expression values for each sample relative to that of the corresponding bridge channel. After combining the data from the runs, we performed a log2 transformation of the expression values, filtered out proteins with too many missing values, and applied VST normalization, all using the DEP v1.16.0 R package[18]. Again, the mean-variance relationship and distribution was checked (Supplementary Figure 1e,f, Supplementary Figure 3e,f). We then imputed missing data using K-nearest neighbor imputation, which did not significantly affect the distribution (Supplementary Figure 4a,c). Finally, we removed any remaining batch effects between the two runs that were amplified through the preprocessing steps using methods from the limma v3.50.3 R package, which are appropriate to apply on properly transformed proteomics data[19]. The removal of batch effects was confirmed using paired PCA plots of the top principal components (PCs) (Supplementary Figure 4b,d). The preprocessed dataset contained 1,786 proteins.

*Bisulfite sequencing data (RR-1 - SOL and TA muscle - female)*

Raw bisulfite sequencing FASTQ files were processed using the Nextflow nf-core methylseq pipeline (v1.6.1), which uses the Bismark aligner for genome alignment and extracting methylation calls (Supplementary Figure 2a)[20,21]. The processed data were filtered to only CpG-type methylated sites, as non-CpG methylation is usually restricted to a few specific cell types (e.g. pluripotent stem cells, glial cells, neurons) that are not as relevant in this context (Supplementary Figure 2b-d)[22]. The methylation sites were then mapped to their corresponding genes. For each methylation site, we found the gene whose chromosomal range included the site. Methylation sites that didn't fall into annotated gene regions were discarded, as the current study focuses on mechanistic relationships between coding features. We then calculated the percentage of CpG sites in each gene that was methylated (% methylation). There were 48,368 and 47,660 methylation features for OSD-104 and OSD-105, respectively, after preprocessing (distributions shown in Supplementary Figure 4e,f). We experimented with using site-level methylation features instead of gene-level methylation features, but this resulted in severe overfitting of machine learning models.

*Calcium reuptake data (RR-1 SOL female, RR-9 TA male)*

Calcium reuptake data was acquired from OSDR dataset OSD-488[15], which contains rates of Ca2+ uptake in the muscle homogenates measured in a 96-well plate using the Indo-1 Ca2+ fluorophore[1]. These values were collected as a time-series, with the measurements of cytoplasmic calcium concentrations taken at multiple points in time during a period of muscle relaxation. For our analysis, we use area under the curve (AUC) as a measurement of calcium reuptake change over time. A lower AUC value implies more efficient calcium reuptake.

For this study, we did not have multi-omics and calcium uptake measurements from the same animals. Therefore, for comparing omics and calcium uptake data, we assigned calcium reuptake values to omics samples based on perturbation analysis to identify the optimal pairing (see Supplementary Information, Supplementary Figure 7). It is well characterized that there are significant physiological and molecular differences between muscle samples from spaceflight and ground samples[1,2]. We inferred that this difference would be greater than within-group differences in mission, age, and sex, and would allow us to identify spaceflight effect relationship between omics features and calcium uptake. Specifically, the SOL calcium reuptake measurements were collected from age- and sex-matched mice from the same RR-1 cohort as the OSD-104 SOL omics data. There was no TA calcium reuptake measurement done in the RR-1 cohort; OSD-488 dataset only had TA calcium reuptake measurements collected from 10-week-old male mice flown on the RR-9 mission. We therefore paired the TA muscles calcium reuptake from these 10-week-old male mice with the OSD-105 TA multi-omics data which are from older females.

*Model hyperparameters*

The primary hyperparameters for QLattice (*feyn* package v3.0.2) were the number of epochs and the maximum complexity of the architectures. The number of epochs corresponded to the number of generations for the evolutionary search algorithm as a whole rather than the number of epochs of backpropagation for any individual model architecture being explored. We tried various values for the number of epochs between 10 and 100, but there were no significant differences in validation performance nor feature rankings. The maximum architectural complexity was restricted to 4 (2 features and 2 functional interactions) for the SOL analysis, since SOL data had two data types and we were interested in modeling the interactions between the data types. Similarly, the maximum architectural complexity was restricted to 6 for the TA analysis.

*Statistical Analysis*

Gene set enrichment analysis was performed using the Enrichr implementation in the *gseapy* library (v0.10.4) in Python, using GO_Biological_Process_2021 as the background gene set. Boxplots were generated using *seaborn* (v0.11.2) in Python, with statistical annotations calculated using the *statannotations* package (v0.5.0) implementation of Mann-Whitney-Wilcoxon two-sided test.

# Results

*Multi-omics biomarkers associated with spaceflight calcium reuptake aberrations are revealed by machine learning regression analysis*

Exposure to spaceflight has been previously reported to increase $Ca^{2+}$ uptake in mouse TA muscles and decrease $Ca^{2+}$ uptake in mouse SOL muscles[1]. We hypothesized that this phenotypic change can be further understood by examining the relationships between genes, proteins, and methylation markers. Therefore, we trained QLattice to identify multi-omics biomarkers predictive of changes in calcium reuptake capacity, using multi-omics datasets from OSD-104 mouse SOL muscle and OSD-105 mouse TA muscle, from female C57BL/6J mice flown at 16-weeks of age on the RR-1 mission. For each muscle type, multi-omics data were combined and subject to dimensionality reduction prior to QLattice training and calcium reuptake levels were used as a target (see Methods and Supplementary Information). We matched RR-1 SOL multi-omics data with RR-1 SOL calcium data, and RR-1 TA multi-omics data with RR-9 TA calcium data. In the latter case, note that RR-9 includes 10-week male mice while RR-1 includes 16-week female mice, but we hypothesized that the spaceflight muscle effect would be great enough to overcome these differences. More details are provided in the Methods and Supplementary Information.

In the TA regression analysis reported here (Figure 1), we used both RNA-seq and proteomics data. We excluded the methylation data because including it reduced QLattice performance while producing very similar results (see Supplementary Information). The top two features predictive of $Ca^{2+}$ uptake rate were Acyp1 and Rps7 proteins (Figure 1b). Representative models containing Acyp1 and Rps7 are shown in Figure 1a. These models consisted of bivariate gaussian functions, bivariate multiply functions, and univariate tanh functions (see Supplementary Information for additional discussion of model architectures). Of the 27 models containing Rps7, 24 included a relationship with Acyp1, possibly indicating a biological interaction between the two features that could be tested in the future through laboratory studies. Gene set enrichment analysis revealed significant enrichment of biological signaling involved in apoptosis, endocytosis, and protein localization (Figure 1d).

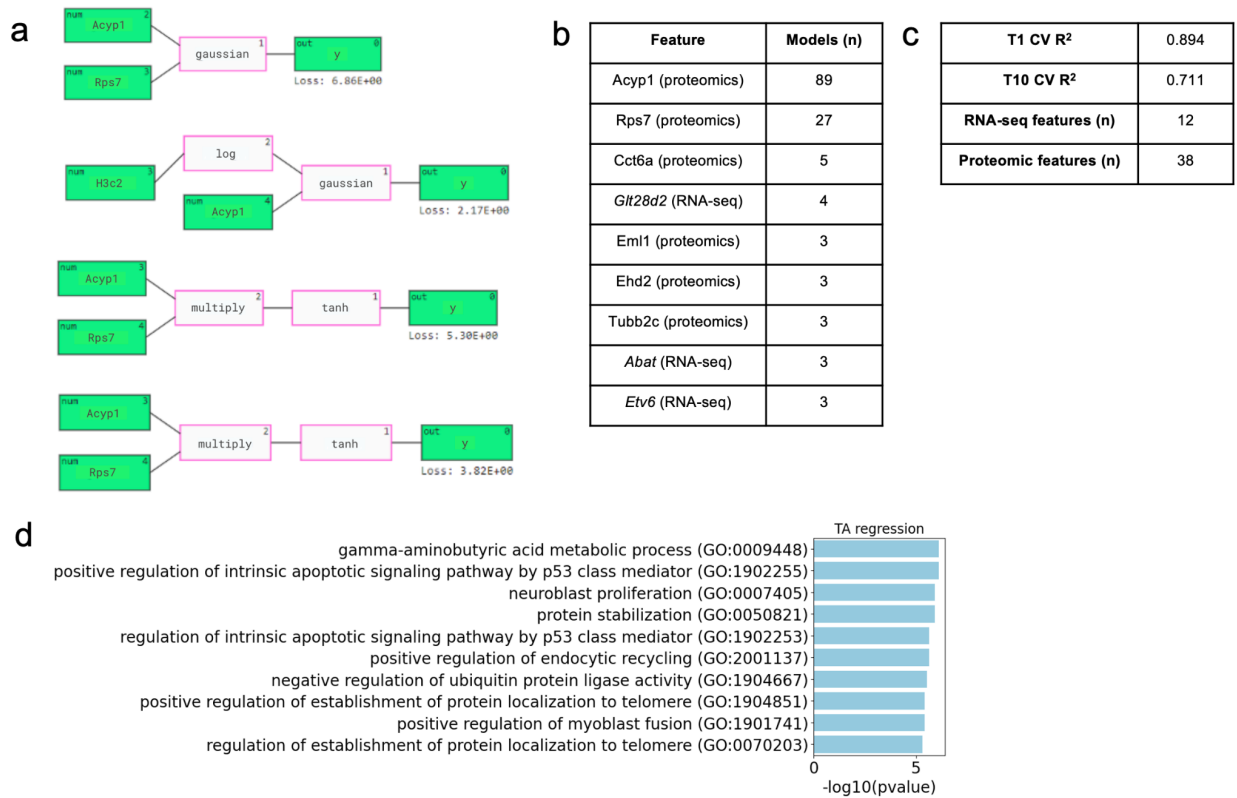

**Figure 1. QLattice regression analysis of TA multi-omics data and calcium uptake. a.** Representative examples of the mathematical relationships between multi-omic features identified by QLattice to predict calcium reuptake in TA muscle during LOOCV. **b.** Top 9 features ranked by how many times they were used in a model found by QLattice during LOOCV. **c.** T1 and T10 cross-validated $R^2$ scores, as well as the number of RNA-seq and proteomic features that were found among the top 50 features. **d.** Gene set enrichment analysis results using the top 9 genes from the QLattice analysis.

In the SOL regression analysis, the best performing models mainly displayed relationships between expression of different genes on the RNA level, related by mathematical functions such as gaussian, linear, and exp (Figure 2a-c). We focused on the top 13 features across all models by rank, all of which were RNA-seq features: *Gm35576, Rspo3, Gpc4, Klhl31, Sox6, Auts2, Sobp, Mdga1, Aox1, Tle4, Klhl33, Eepd1, Rhbdl3,* and *Gm21955*. Gene set enrichment analysis revealed significant enrichment of biological signaling involved in cellular differentiation, synapse organization and assembly, and neuron migration (Figure 2d).

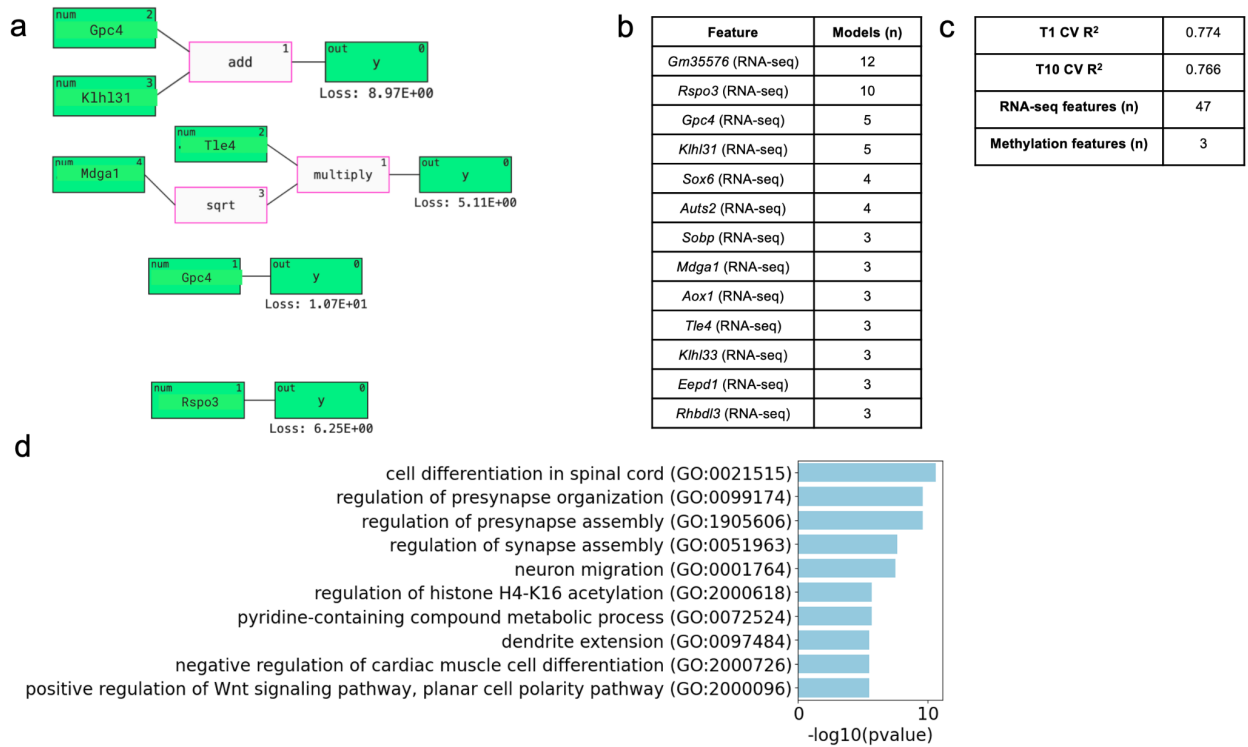

**Figure 2. QLattice regression analysis of SOL multi-omics data and calcium uptake. A)** Representative examples of the mathematical relationships between multi-omic features identified by QLattice to predict calcium reuptake in SOL muscle during LOOCV. **B)** Top 13 features ranked by how many times they were used in a model found by QLattice during LOOCV. **C)** T1 and T10 cross-validated $R^2$ scores, as well as the number of RNA-seq and methylation features that were found among the top 50 features. **D)** Gene set enrichment analysis results using the top 13 genes from the QLattice analysis.

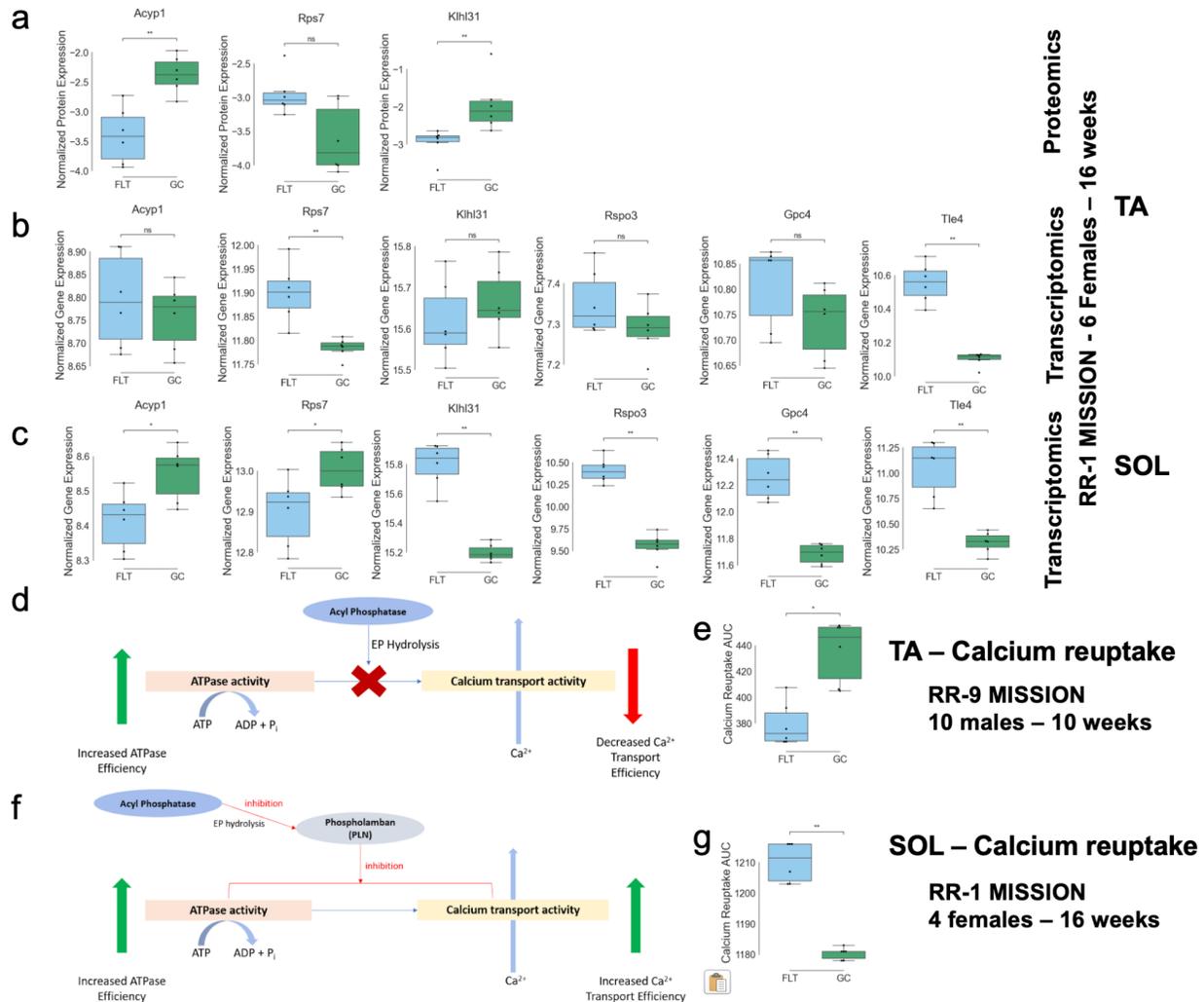

**Figure 3. Gene/Protein relationship with calcium reuptake in SOL and TA muscles and putative mechanism.** The expression levels of the top key genes identified by QLattice analysis and their corresponding protein levels are shown against the calcium reuptake AUC in both TA and SOL muscles flown in space (FLT) or from ground controls (GC) **a)** QLattice key protein levels in TA. **b)** QLattice key gene expression levels in TA. **c)** QLattice key gene expression levels in SOL. **d)** Putative mechanism based on the Acyp1 response, showing up in the majority of the models for TA muscle and **e)** SOL muscle. **f)** Calcium reuptake AUC in TA muscle. **g)** Calcium reuptake AUC in SOL muscle. All significance calculated using Mann-Whitney-Wilcoxon test two-sided: *: $1.00e-02 < p <= 5.00e-02$, **: $1.00e-03 < p <= 1.00e-02$).

QLattice analysis identified 9 and 13 critical genes playing an important role in calcium reuptake of TA and SOL muscles respectively, in the context of spaceflight inducing muscle loss. Figure 3 depicts the changes of expression level in FLT versus GC for some of these genes and their corresponding proteins (when measurements are available) in both muscle types. Many of the genes identified have been described in the literature as playing a role in muscle recovery and we focus our attention on these genes.

In the case of TA muscle, Acyp1 protein was found in 89 models. Acyp1 has been shown to inhibit the activity of Ca2+ transporters in non-phospholamban-associated calcium-dependent

ATPases, such as SERCA-1, which is predominantly found in fast-twitch muscle fibers, the dominant fiber type in the TA muscle[23–25]. In line with this, Acyp1 protein expression is negatively correlated with calcium reuptake in QLattice models for TA muscles (Supplementary Figure 8), indicating that high levels of Acyp1 are associated with low calcium reuptake efficiency AUC. Consistent with previous studies[1], FLT TA samples exhibit lower Acyp1 protein expression and improved calcium reuptake (Figure 3a,e). Interestingly, *Acyp1* gene expression in TA has a much greater spread across samples resulting in no significant difference between FLT and GC (Figure 3b), possibly indicating a more reliable measurement from proteomics than RNA-seq.

Furthermore, Acyp1 has been shown to enhance the activity of Ca2+ transporters in phospholamban-associated calcium-ATPases, including SERCA-2a, which is predominantly found in slow-twitch fibers, the dominant fiber type in SOL muscle. We examined whether this pattern is consistent in the OSD-104 SOL multi-omics data and discovered that *Acyp1* gene expression was also downregulated in SOL FLT relative to GC (Figure 3c), while calcium reuptake was impaired in SOL as previously reported[1] (Figure 3g). Note that QLattice was not trained to find this association for SOL muscle. Such finding is therefore quite strong suggesting a potential mechanism. The observed down regulation in *Acyp1* could contribute to impairments in SERCA function found in the FLT SOL, as it is known that phospholamban is highly expressed in this muscle and is less expressed in fast glycolytic muscles. Taken together, these findings suggest that Acyp1 may play a mechanistic role in the dysregulation of calcium induced by spaceflight (Figure 3d,f), but additional research is necessary to establish this relationship .

The second key protein identified in TA muscle was Rps7. This gene is known to be downregulated by nitrosative stress[26], which is related to impaired calcium reuptake[1]. Consistent with this, Rps7 is positively correlated with calcium reuptake capacity efficiency in the QLattice models (negatively associated with calcium reuptake AUC; Supplementary Figure 8). This suggests a potential role for Rps7 in the calcium reuptake response to nitrosative stress. No nitrosative stress was reported in the TA samples from the original study[1], suggesting that mechanisms possibly including Rps7 may have enhanced the calcium reuptake efficiency. Further studies would be required to establish these relationships.

The QLattice analysis for SOL revealed a very distinct set of genes (Figure 2b). This is however not surprising as only gene expression and methylation data were available. Several of these genes are already known for their relationship to calcium reuptake efficiency and muscle response to injury and their levels are plotted for protein expression (when available) and gene expression in both TA and SOL muscle in Figure 3a-c. *Gpc4* is underexpressed in injury-activated muscle satellite cells[27]; similarly, *Tle4* is normally underexpressed following muscle injury to allow myogenesis[28]. In our data, both *Tle4* and *Gpc4* were upregulated in mouse FLT SOL (Figure 3a-b), which displayed impaired calcium reuptake vs. GC SOL (Figure 3g). This may indicate that failure of proper *Gpc4* and *Tle4* downregulation may play a role in damaged calcium reuptake, possibly by lowering overall muscle quality. This is supported by previous RNA-seq data showing that genes involved with myogenesis and differentiation were downregulated in the FLT SOL from mice[29]. Alternatively, *Tle4* expression is known to be triggered by calcium signaling[30], so the observed *Tle4* upregulation may instead be the result of increased cytoplasmic calcium levels due to impaired reuptake.

Further, *Rspo3* has been found to be one of the most upregulated genes after SOL training and is associated with decrease in muscle atrophy[31], and its knockout has shown to compromise myogenesis and myotube differentiation[32]. Similarly, mice lacking *Klhl31* exhibit stunted skeletal muscle growth, centronuclear myopathy, and SR dilation[33]. In our data, both *Rspo3* and *Klhl31* are also upregulated in SOL FLT samples with lower calcium reuptake ability (Figure 3c), possibly as a compensatory or adaptive mechanism to increased calcium levels due to decreased uptake[34].

*Multi-omics biomarkers associated with spaceflight calcium reuptake aberrations are revealed by machine learning classification analysis*

We then hypothesized that there may be other molecular pathways affected by spaceflight in mouse muscle that could be identified through feature relationships in QLattice models. Therefore, we broadened the scope of our analysis to identify multi-omics features that were predictive of the FLT or GC groups, rather than restricting to a single phenotype. We used QLattice to classify FLT samples from GC and assessed the resulting models and feature interactions.

In the TA classification analysis, we used all 3 types of omics data: RNA-seq, proteomics, and methylation data (Figure 4a-c). The top 11 features from this analysis included all 3 types of features: *Trak2* (RNA-seq), *Tle4* (RNA-seq), *Tspan4* (RNA-seq), Actin (Proteomic), Gm22281 (Methylation), *Sell* (RNA-seq), Ech1 (Methylation), *Fhod1* (RNA-seq), *Egr2* (RNA-seq), *Klhl21* (RNA-seq), and *Lrp2bp* (RNA-seq).

In keeping with our hypothesis, gene set enrichment analysis revealed significant enrichment of pathways relevant to muscle biology and the neuromuscular response to stress, including skeletal muscle cell differentiation, positive regulation of myelination, and Schwann cell differentiation (Figure 4d). Interestingly, multiple pathways involved mitochondrial regulation, which has been previously identified as a molecular response to spaceflight in multiple tissues including muscle[35].

The pathway analysis also uncovered perturbation of actin and myosin structural regulation. The actin protein was the top proteomics feature found across QLattice models (Figure 4b) and was upregulated in TA FLT samples (Figure 5a). Actin is a key component in the myofibril bundles which generate muscle contractions after $Ca^{2+}$ release and signaling[36]. Further, the top RNA-seq feature *Trak2* is known to enable myosin binding activity for muscle contraction[37]. *Trak2* is also involved in Rho GTPase cycle which plays an important role in muscle mass regeneration and myofibrillogenesis[37]. The *Trak2* gene is upregulated in FLT TA muscle in our data (Figure 5b), possibly as a muscle regeneration mechanism in a weightless environment. *Tle4*, the second highest occurring RNA-seq feature across all QLattice models, acts as a corepressor regulating muscle cell differentiation[28]. The *Tle4* gene is upregulated in FLT TA muscle in our data (Figure 5b), possibly due to the lack of a need for skeletal muscle growth in a weightless environment.

Interestingly, both *Trak2* and *Tle4* displayed some co-occurrence with Actin. Out of the 20 models that *Trak2* appeared in, 4 of them contained Actin, while out of the 14 models containing *Tle4*, 2 of them contained Actin. This may indicate a co-regulation network between Actin structural muscle activity and muscle mass regeneration and cell differentiation in response to spaceflight, which could be further investigated in laboratory studies.

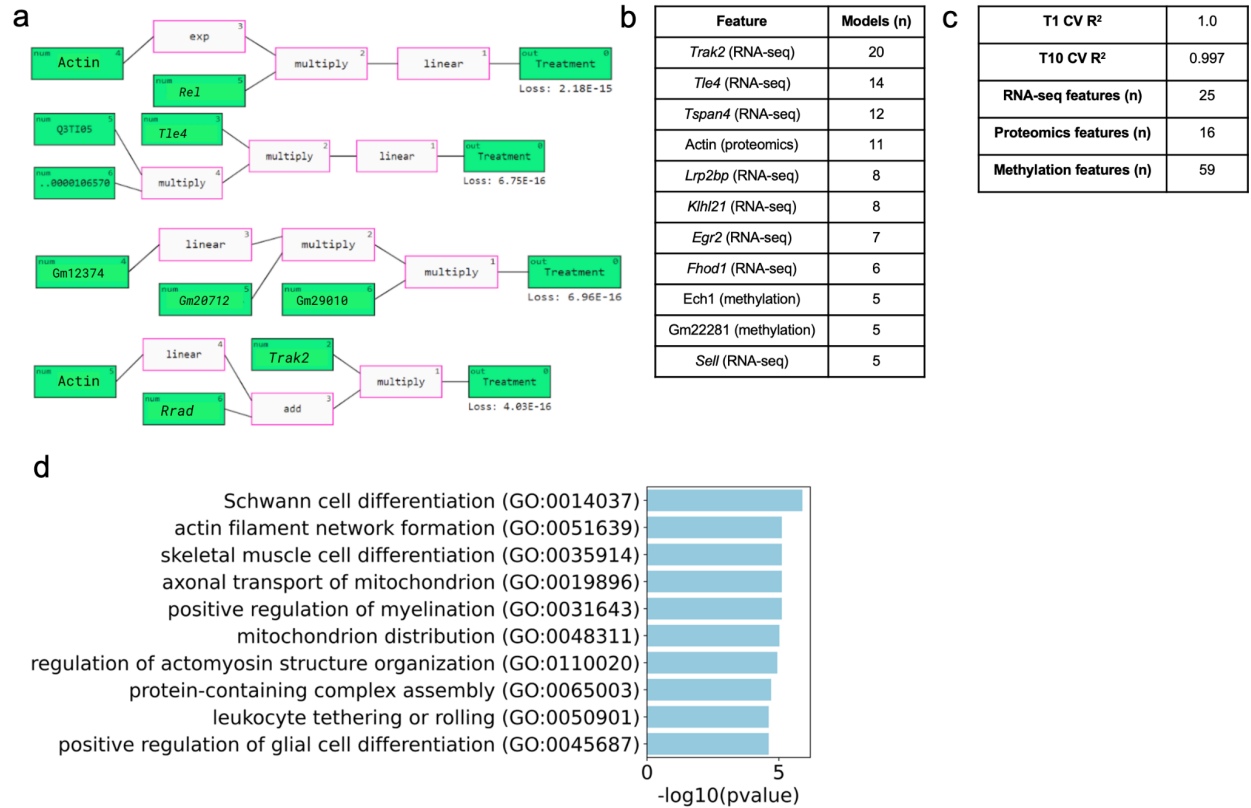

**Figure 4. QLattice classification analysis of TA multi-omics data and FLT/GC groups. a)** Representative examples of the mathematical relationships between multi-omic features identified by QLattice to predict FLT versus GC in TA muscle during LOOCV. **b)** Top 11 features ranked by how many times they were used in a model found by QLattice during LOOCV. **c)** T1 and T10 cross-validated $R^2$ scores, as well as the total number of RNA-seq, proteomics and methylation features across all models. **d)** Gene set enrichment analysis results using the top 11 features from the QLattice analysis.

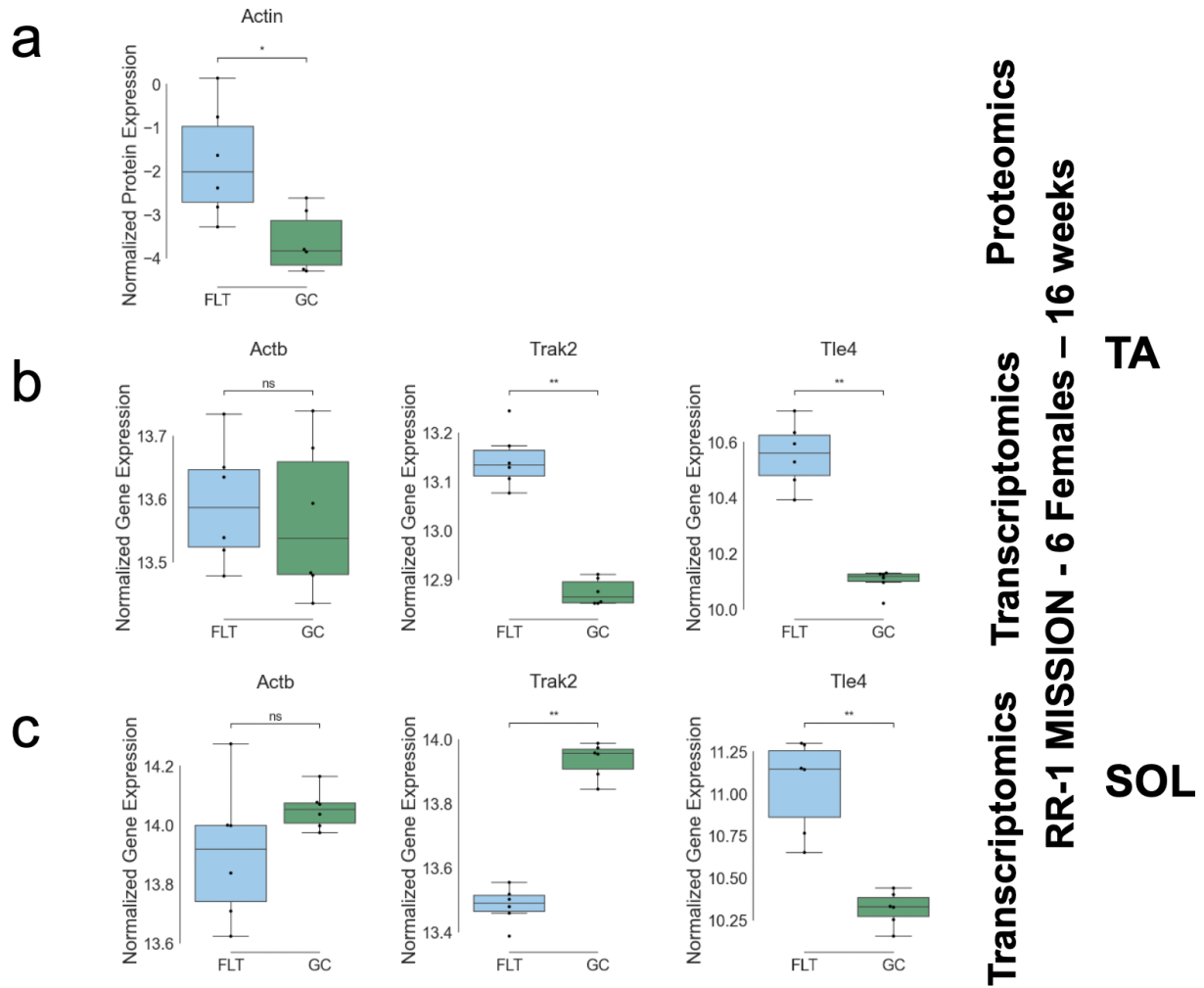

**Figure 5. Expression levels of top proteins and genes identified by QLattice TA and SOL classification analysis and muscle weights. a)** QLattice key protein levels in TA. **b)** QLattice key gene levels in TA. **c)** QLattice key gene levels in SOL. All significance calculated using Mann-Whitney-Wilcoxon test two-sided: *: 1.00e-02 < p <= 5.00e-02, **:1.00e-03 < p <= 1.00e-02).

In the SOL classification analysis, RNA-seq gene features comprised 69 out of the 80 features across all resulting models (Figure 6a-c). The top 9 recurrent features were *Fam220a*, *Lrp4*, *Osgin2*, *Gm29686*, *Gm22281* (Methylation), *Sema6c*, *Alpk3*, *Tmod1*, and *Bcam*. For the most part, these features appeared in single-feature models, related to the FLT/GC outcome by a linear, log, or inverse relationship. Of the 120 total models, 18 described relationships between 2 features. Interestingly, 11 of these were pairs of methylation and RNA-seq features, indicating a potential cooperative relationship between gene expression and DNA methylation in spaceflight SOL muscle response.

Similar to the calcium reuptake prediction analysis, the SOL FLT/GC classification analysis mainly identified models with interactions between RNA-seq gene features. Gene set enrichment analysis of the top 11 features revealed enrichment of pre- and post-synaptic membrane assembly and organization (Figure 6d), in keeping with previous research showing structural alterations in muscle synaptic organization in spaceflight[38].

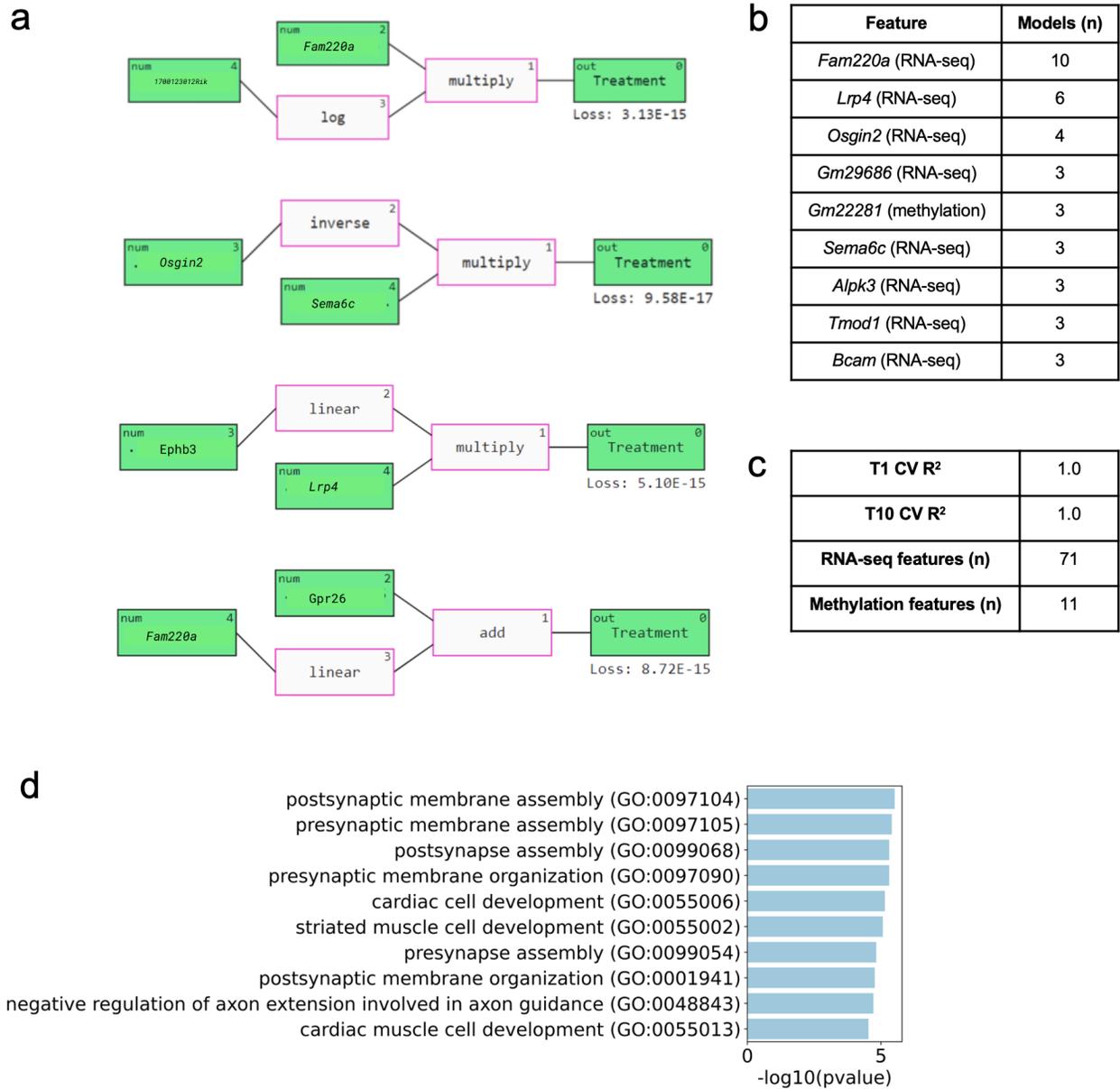

**Figure 6. QLattice classification analysis of SOL multi-omics data and FLT/GC groups. a)** Representative examples of the mathematical relationships between multi-omic features identified by QLattice to predict FLT versus GC in SOL muscle during LOOCV. **b)** Top 9 features ranked by how many times they were used in a model found by QLattice during LOOCV. **c)** T1 and T10 cross-validated $R^2$ scores, as well as the total number of RNA-seq and methylation features across all models. **d)** Gene set enrichment analysis results using the top 9 features from the QLattice analysis.

Discussion

Here we report the identification of multi-omics biomarkers in mouse SOL or TA muscle that are predictive of change in calcium reuptake capacity during spaceflight, or broadly predictive of molecular changes in spaceflight samples compared to ground control. Our study is a novel contribution to the field both in the identification of these biomarkers, and in that we demonstrate the utility of ML methodology for biomarker discovery in space biology by using the QLattice symbolic regression method to characterize biomarker relationships to each other and to the predictive target. Instead of using the traditional system biology approach, ML methods provide unbiased identification of potential candidates for biological mechanisms. Our study is also one of the first to analyze combined multi-omics and non-omics/phenotypic data from the NASA OSDR, leveraging the relational database structure which allows for mapping samples across assays and missions.

Appropriate regulation of muscular cytoplasmic calcium levels are key for several downstream calcium-dependent signaling pathways. Calcium reuptake is reportedly improved in TA muscle, but impaired in SOL muscle, during spaceflight[1], with limited understanding of the molecular signaling causing these phenotypic changes. Here, we report that enhanced $Ca^{2+}$ uptake in FLT TA is directly related to the combined interaction of Acyp1 protein downregulation and Rps7 protein upregulation; while decreased $Ca^{2+}$ uptake in FLT SOL is related to interactions of upregulation of several different pairs of genes including *Gpc4*, *Tle4*, *Rspo3*, and *Klhl31*. The lower $Ca^{2+}$ uptake in FLT SOL is also correlated with significant weight loss for the SOL muscle in the RR1 flight samples (t-test pvalue < 0.05); whereas there is no significant change of weight for TA muscles (Table 1). Overall, the analysis provided here suggests that TA muscles are more resilient to space conditions and Acyp1 and Rps7 seem to be good candidates to counteract weight losses and poor $Ca^{2+}$ uptake observed in SOL muscles.

|  |  | Average weight (mg) |
|---|---|---|
| RR-1 SOL | FLT | 7.9 |
|  | GC | 10.5 |
| RR-1 TA | FLT | 13.3 |
|  | GC | 13.9 |

**Table 1. Average processed tissue weights for RR-1 SOL and TA groups.**

On the technological aspect of this work, we show that one of the major advantages of QLattice compared to traditional multi-omics or differential gene expression analysis methods is its ability to elucidate a variety of mathematical interactions, not necessarily linear, between different multi-omics features. We noted a variety of mathematical functions in the QLattice models from our analysis, including bivariate gaussian, bivariate multiply, univariate tanh, addition, multiplication, and log. The response plots provided by QLattice help translate the mathematical functions into biological and mechanistic interpretations (Supplementary Figure 8). We noted that when gaussian functions are reported, only the first half of the gaussian curve is being used to fit a sigmoidal trend (Supplementary Figure 8). Biologically, we interpret this to be because biological quantities (e.g. calcium concentration) cannot be negative, nor cannot exceed a certain threshold (e.g. the total amount of calcium stored in the sarcoplasmic reticulum). Thus, linear models would actually fail to extrapolate in these more extreme ends of the quantity's range, while sigmoidal models would fit much more accurately. To further improve the predictive capability of this tool, it would be valuable to further characterize the relationship between the mathematical functions identified by QLattice and the distribution of protein concentrations in various tissues. Such characteristics would help further translate the interactions of the various functions found in QLattice with an actual biological process between different key components of a tissue.

While additional study is needed to fully characterize the relationships identified here by QLattice, we suggest that QLattice's emphasis on concise, interpretable models makes it an especially appropriate ML methodology for research applications, where explainability is key. Biomedical research, and especially space biology research, presents the additional challenge of small sample sizes, which usually leads to severe overfitting and lack of generalizability in the models found by conventional ML algorithms. This issue is at least partially addressed by QLattice's predisposition to limit the architectural complexity of models (which has a strong regularizing effect), as well as its heavy intrinsic feature selection based on mutual information with the target variable.

As demonstrated in this study, when a ML method properly accounts for these challenges and priorities specific to space biology research, it can provide significant guidance for what to focus on in future research. Although the top biomarkers from QLattice must be further characterized and confirmed, the method has appreciably narrowed the research "search space" by directing us toward groups of biomarkers that are most promising. In this way, explainable and interpretable ML, with its advantage over humans in being able to process huge feature spaces, serves as a metaphorical "metal detector," telling us where we should start digging.

We conclude with our observations on the contributions of the different types of omics data to the QLattice predictive models. In this work, DNA methylation CpG+ features, when mapped to gene names or when maintained as genomic coordinates, failed to greatly contribute to model architectures and resulted in lower predictive performance. We suggest that this may be because methylation marks are deposited over time, so molecular changes during spaceflight are primarily dominated by functional changes while small but persistent epigenetic changes may be better captured upon return to earth. To test this hypothesis, future studies could capture methylation measurements both before and after spaceflight from the same animals. In accordance with this hypothesis, both protein and gene expression changes were the most predictive features, with the top proteomic features much stronger and more cohesive than those of top RNA-seq features. This may constitute support in favor of focusing on proteomic analysis over RNA-seq analysis in future spaceflight studies, as the relationship between proteins and function is more immediate compared to the presumably noisier relationship between transcripts and function.

Taken together, our results build on previous work in the field by reporting a promising demonstration of an explainable ML method for space biology research, and providing several potential biomarkers for future study on muscle response to spaceflight.

## Author Contributions

## Competing Interests Statement

## Acknowledgements

Resources supporting this work were provided by the NASA High-End Computing (HEC) Program through the NASA Center for Climate Simulation (NCCS) at Goddard Space Flight Center. We acknowledge support from the Space Life Sciences Training Program. The Open Science Data Repository is funded by the Space Biology Program (Science Mission Directorate, Biological and Physical Sciences Division) of the National Aeronautics and Space Administration. Funding for open access charge: NASA.

## Supplementary Information

*Dimensionality reduction*

Following the preprocessing of the omics datasets, we then applied Multi-Omics Factor Analysis (MOFA) on the combined multi-omic datasets using the MOFA2 v1.4.0 R package[39]. MOFA is an unsupervised dimensionality method that uses Bayesian prior/posterior distribution updating to find a small set of latent factors to capture the variance in the data, which are analogous to principal components in PCA[39]. Unlike PCA, however, MOFA relies on probabilistic principals instead of geometric properties and is able to find its latent factors using information from multiple omics datasets (e.g. a factor is determined based on information provided by the transcriptomic, proteomic, and epigenomic datasets). Additionally, MOFA assumes that the distribution is continuous and roughly Gaussian, which was the case for our normalized and transformed datasets (Supplementary Figure 3, Supplementary Figure 4A,B,E,F).

MOFA Factor 1, which captures most of the variance in the dataset, cleanly separated FLT and GC samples in both datasets (Supplementary Figure 5A,B). We inferred that the features associated with Factor 1 would be most relevant for examining expression differences underlying spaceflight-induced physiological changes; therefore, for each data type, we selected only the top K features associated with MOFA Factor 1 to use for downstream methods.

To derive K, we also performed a Weighted Gene Correlation Network Analysis (WGCNA v1.71)[40] on each data type and extracted features from clusters with significant differences between FLT and GC samples as follows. In WGCNA, all features within each dataset were sorted into clusters based on their respective correlation between FLT and GC groups. The clusters were sorted by correlation, and the features contained in the clusters with the highest correlation (>0.18) were selected. This cutoff was selected so that a sufficient number of features could be selected from each dataset to allow for an ample overlap with MOFA selection. We then chose K to give maximum % overlap between MOFA Factor 1 features and WGCNA features (Supplementary Figure 6), since overlap between important features extracted from two different unsupervised methods increases confidence in the true significance of those features.

After evaluation, we found that the factors found by MOFA were primarily driven by RNA-seq and proteomic features, with very few methylation features (Supplementary Figure 5C,D). Therefore, for methylation feature reduction, we ran MOFA with only the methylation dataset. There were no factors that clearly separated the FLT vs GC, so we selected top methylation features associated with all factors found by MOFA (5 total) whose variance contributions were not driven by outliers (one factor was not included because outlier data points drastically inflated its variance contribution)[7]. WGCNA was finally used to identify methylation features from the top MOFA features that gave highest correlation between GC and FLT.

When feeding features from different omics types into QLattice, all features were automatically set to comparable scales within QLattice using a variation of min-max scaling [7,41].

*Cross validation*

Due to the small sample size of our dataset, we decided to perform leave-one-out cross validation (LOOCV). During each iteration of LOOCV, the model is trained on the entire dataset except one sample, which is used as a validation sample to test the model's generalizability. This is done for *n* iterations (where *n* is the size of the dataset) so that each sample is used as a validation sample at some point. Because the training set for each iteration of LOOCV contains only one fewer sample than the full dataset, LOOCV is approximately statistically unbiased[42].

Since QLattice outputs the top 10 model architectures for each run, the 12 iterations of LOOCV generated 120 distinct models, where each was used to predict either the calcium reuptake AUC label (in the regression problem) or the FLT/GC label (in the classification problem) of one validation sample. For the regression problem, we then calculated the multiple $R^2$ between these 120 validation sample predictions and the corresponding calcium reuptake AUCs; for the classification problem, we calculated the accuracy among these 120 predictions. We call these the T10 validation scores. Additionally, we used the same measurements, but only considering the top model from each iteration instead of the top 10 (12 models total instead of 120); we call these the T1 validation scores. While these validation metrics do not represent the generalizability of a particular model architecture, they do represent the overall ability of QLattice to find generalizable models and thus inform the overall generalizability of biological results derived from all models.

*Assignment of Calcium Reuptake Labels*

One caveat of our study is that the calcium reuptake measurements were collected from different mice than the omics data. The SOL calcium reuptake measurements were collected from mice from the same RR-1 cohort as the OSD-104 SOL omics data, while the tibialis anterior calcium reuptake measurements were collected from mice flown on the RR-9 mission. We assessed inter- and intra-condition differences in calcium reuptake AUC for both muscle types, and found that the inter-condition differences between FLT and GC mice were much greater than the intra-condition variation, indicating that AUC levels are likely consistent within a cohort of FLT or GC mice. Therefore, for the purposes of relating multi-omics features to overall calcium reuptake levels, we randomly assigned each mouse in OSD-104 and OSD-105 the calcium reuptake AUC of a mouse with the same biological condition and muscle type from OSD-488.

In this process, we found that different random calcium label assignments affected subsequent QLattice validation performance. We generated a distribution of 60 random label assignments and their corresponding QLattice validation multiple $R^2$ values, and found that one particular labeling assignment for each muscle type resulted in drastically higher QLattice validation performance (Supplementary Figure 7).

Our hypothesis is that each mouse in OSD-104/105 is most physiologically similar to a particular mouse in OSD-488, resulting in an optimal labeling assignment when the calcium reuptake values are assigned to the multi-omics features from the most similar mouse. This hypothesis is supported by the fact that for each muscle type, a single labeling assignment drastically outperformed the others, which suggested that these assignments likely did not outperform just by random chance. Therefore, we chose the optimal labeling assignment for each muscle type for downstream machine learning analyses. The biological relevance of the results (see Results) further increased our confidence that the high validation performance of this labeling assignment was not driven only by spurious correlations.

*Various Multi-Omics Combinations*

In addition to the QLattice runs mentioned in the paper, we also tried running QLattice on various other omics combinations.

For OSD-104 regression analysis, we tried running QLattice using only the methylation data, since RNA-seq features seemed to dominate the models in the multi-omics runs. These methylation-only models performed very poorly (validation $R^2 < 0.15$), again demonstrating the lack of useful information contained within gene-level methylation data for this specific study.

For OSD-105 regression analysis, we performed additional runs of QLattice using all three omes (RNA-seq + proteomics + methylation) and only RNA-seq + methylation data. The RNA-seq + methylation data performed quite poorly (validation $R^2 < 0.4$), signifying the importance of proteomic features for the TA data. Interestingly, the runs using all three omes, while not performing terribly (T1 and T10 validation $R^2$ scores of 0.716 and 0.476, respectively), did noticeably worse than the runs using only RNA-seq and proteomics. A potential explanation for this is that the introduction of noisy gene-level methylation features not only doesn't help, but 1) can exacerbate the risk of overfitting and 2) drastically expands the search space for QLattice to explore (this was also suggested when we needed to increase the number of evolutionary epochs QLattice would explore for in order to even get comparable results to the RNA-seq + proteomic only runs). However, despite the lower performance, Acyp1 and Rps7 still dominated in the resulting models when using all three omes.

# Supplementary Figures

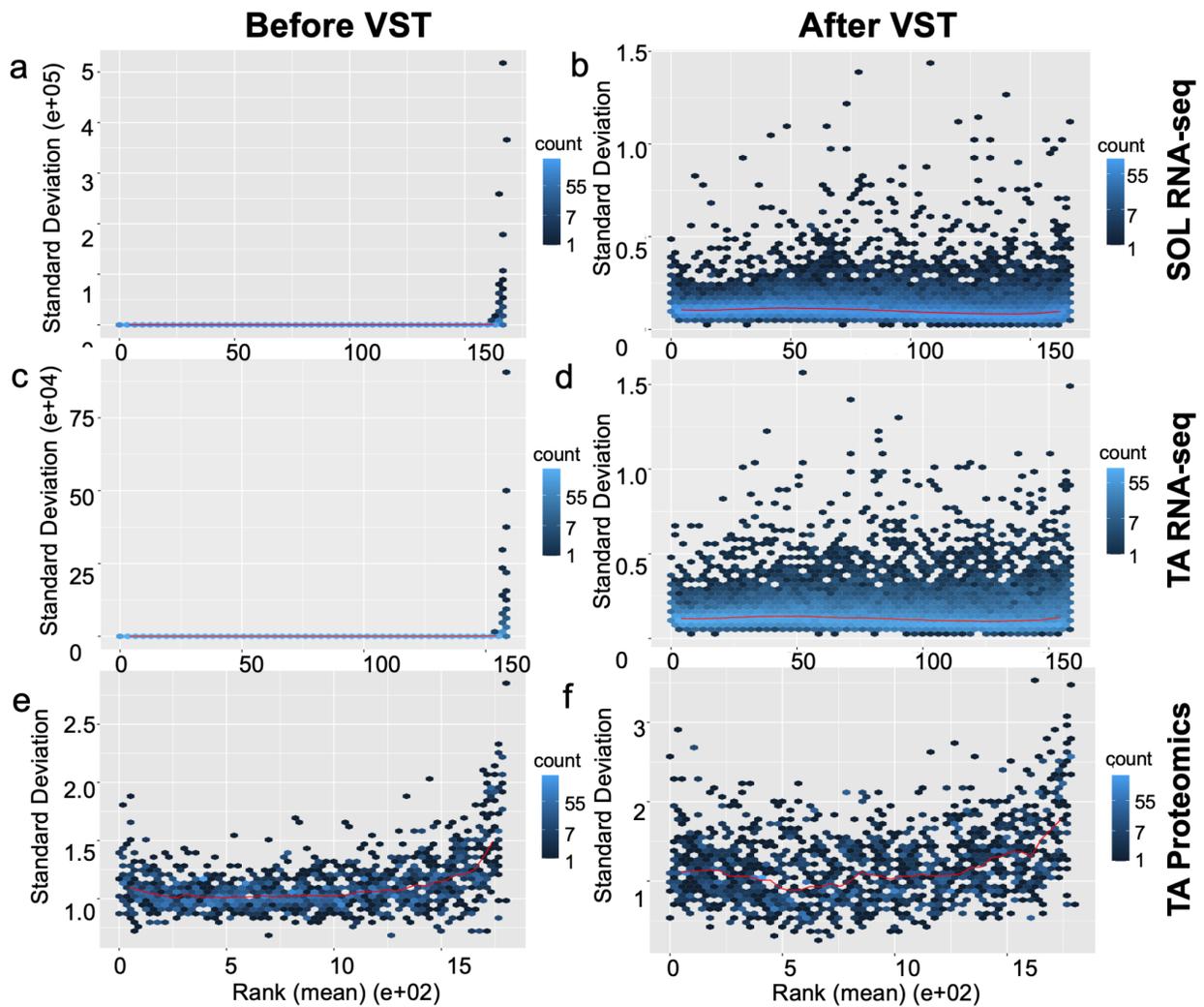

**Supplementary Figure 1. Mean-variance relationships for OSD-104 SOL RNA-seq and OSD-105 RNA-seq and proteomics before and after VST. a)** Mean-variance relationship before VST for SOL RNA-seq. **b)** Mean-variance relationship after VST for SOL RNA-seq. **c)** Mean-variance relationship before VST for TA RNA-seq. **d)** Mean-variance relationship after VST for TA RNA-seq. **e)** Mean-variance relationship before VST for TA proteomics. **f)** Mean-variance relationship after VST for TA proteomics.

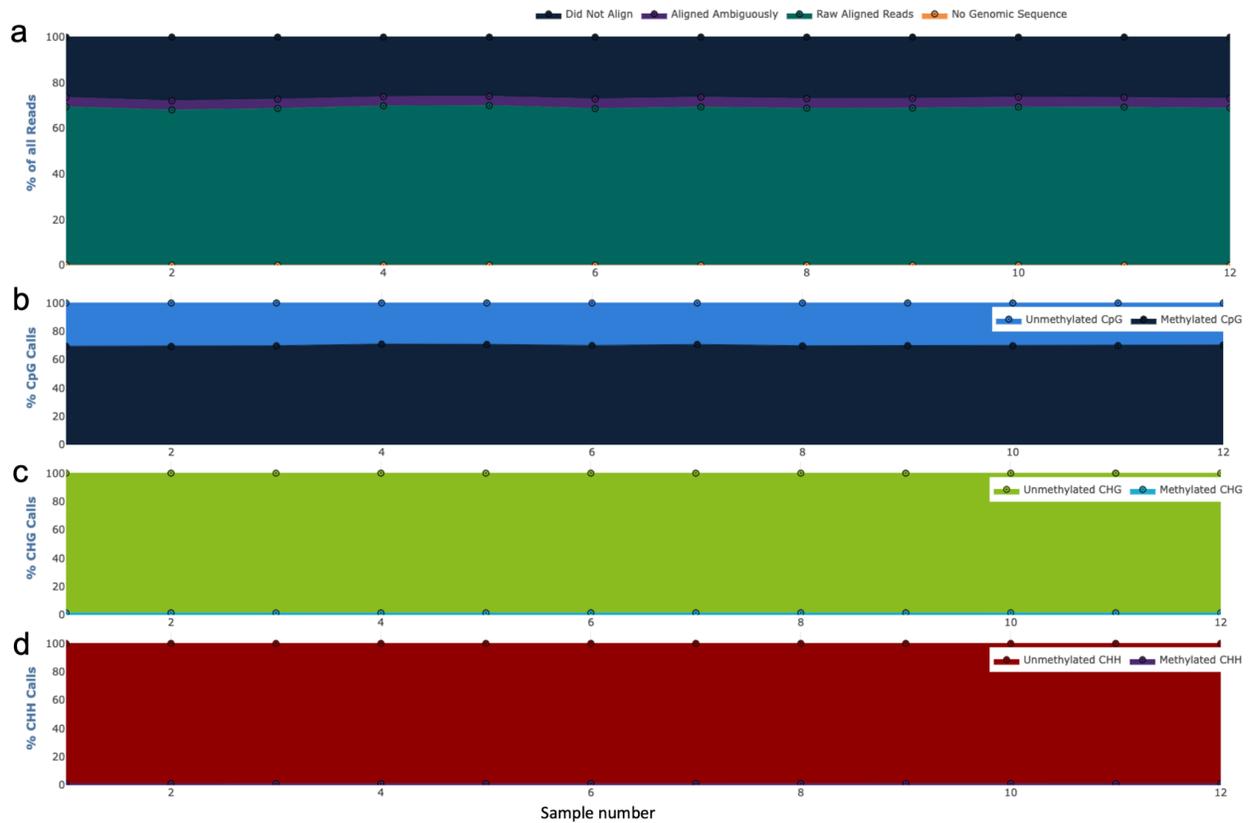

**Supplementary Figure 2. Bisulfite sequencing alignment metrics for OSD-105 TA data. a)** Percent reads aligned for OSD-105 (TA) raw bisulfite sequencing alignment with Bismark. **b)** Percent CpG calls for OSD-105 (TA) bisulfite sequencing data. **c)** Percent CHG calls for OSD-105 (TA) bisulfite sequencing data. **d)** Percent CHH calls for OSD-105 (TA) bisulfite sequencing data.

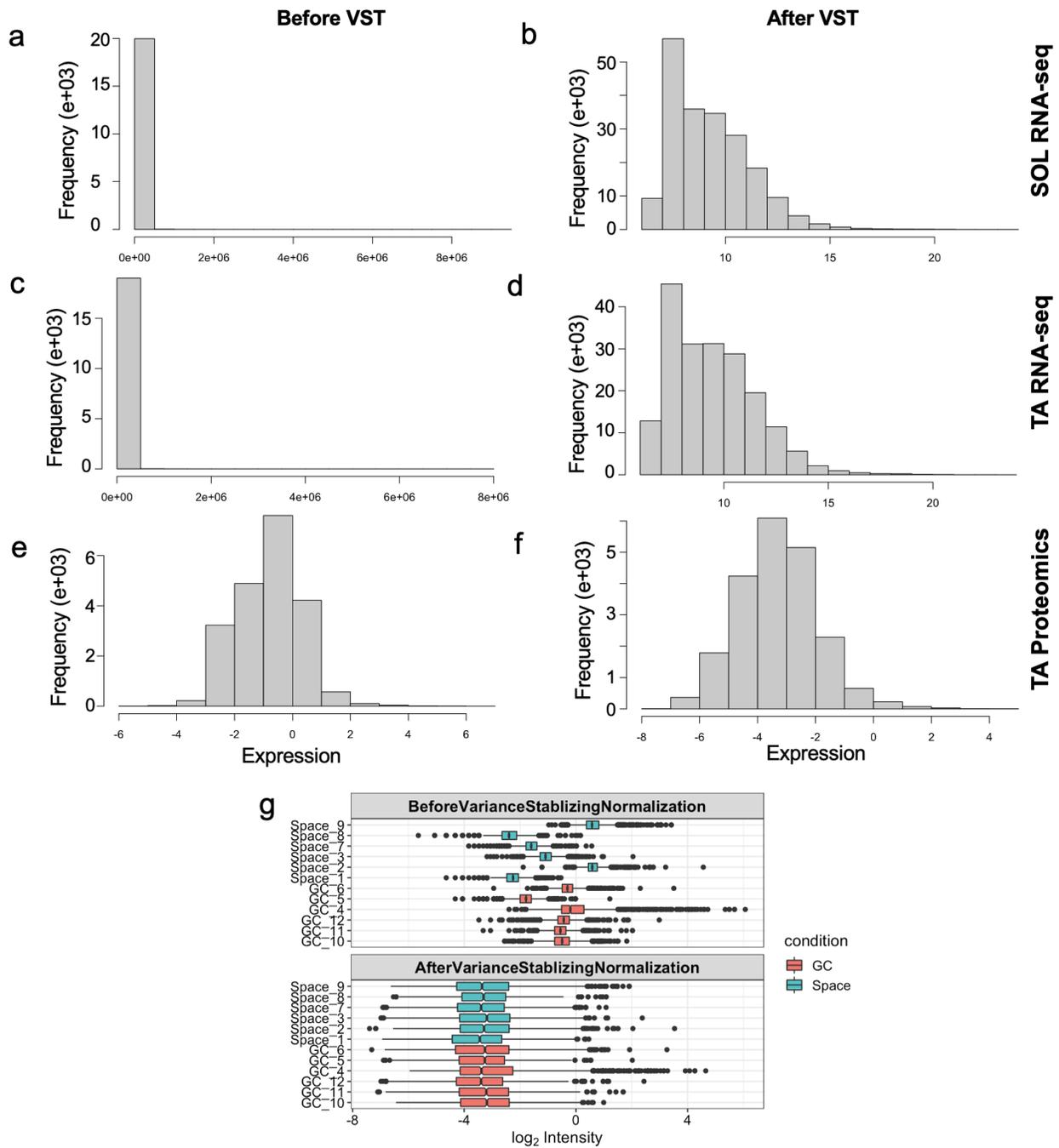

**Supplementary Figure 3. Data distributions before and after VST. a)** Distribution of normalized expression values before VST for SOL RNA-seq. **b)** Distribution of normalized expression values after VST for SOL RNA-seq. **c)** Distribution of normalized expression values before VST for TA RNA-seq. **d)** Distribution of normalized expression values after VST for TA RNA-seq. **e)** Distribution of normalized expression values before VST for TA proteomics. **f)** Distribution of normalized expression values after VST for TA proteomics. **g)** Distributions of normalized expression values for each sample before and after VST for TA proteomics.

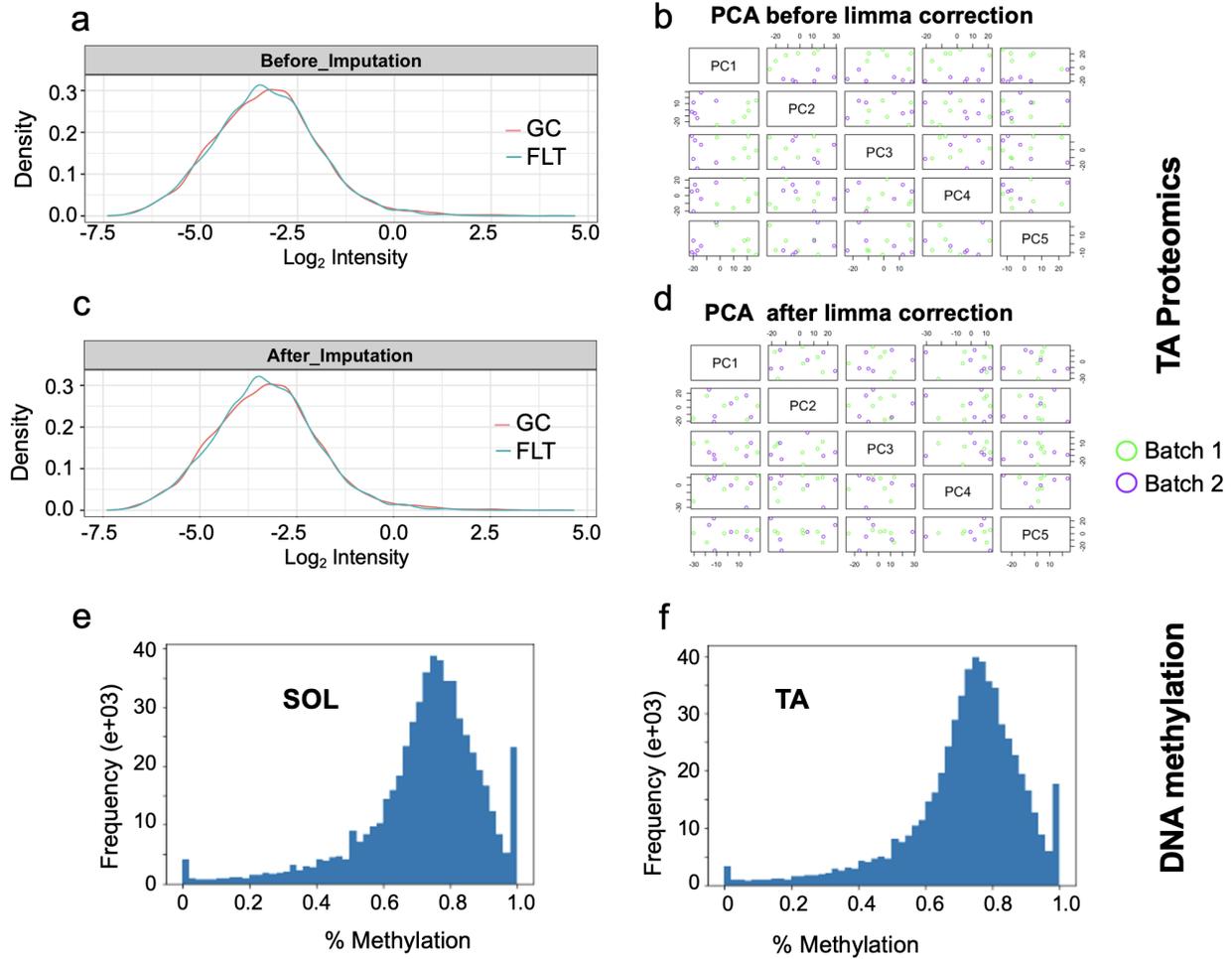

**Supplementary Figure 4. Proteomics data distributions before and after imputation and limma and % methylation distributions. a)** Distribution of TA proteomics data before imputation. **b)** PCA plots for proteomics data color-coded by TMT run before correction for batch effects. **c)** Distribution of TA proteomics data before imputation. **d)** PCA plots for proteomics data color-coded by TMT run after correction for batch effects. **e)** Distribution of % DNA methylation values for SOL. **f)** Distribution of % DNA methylation values for TA.

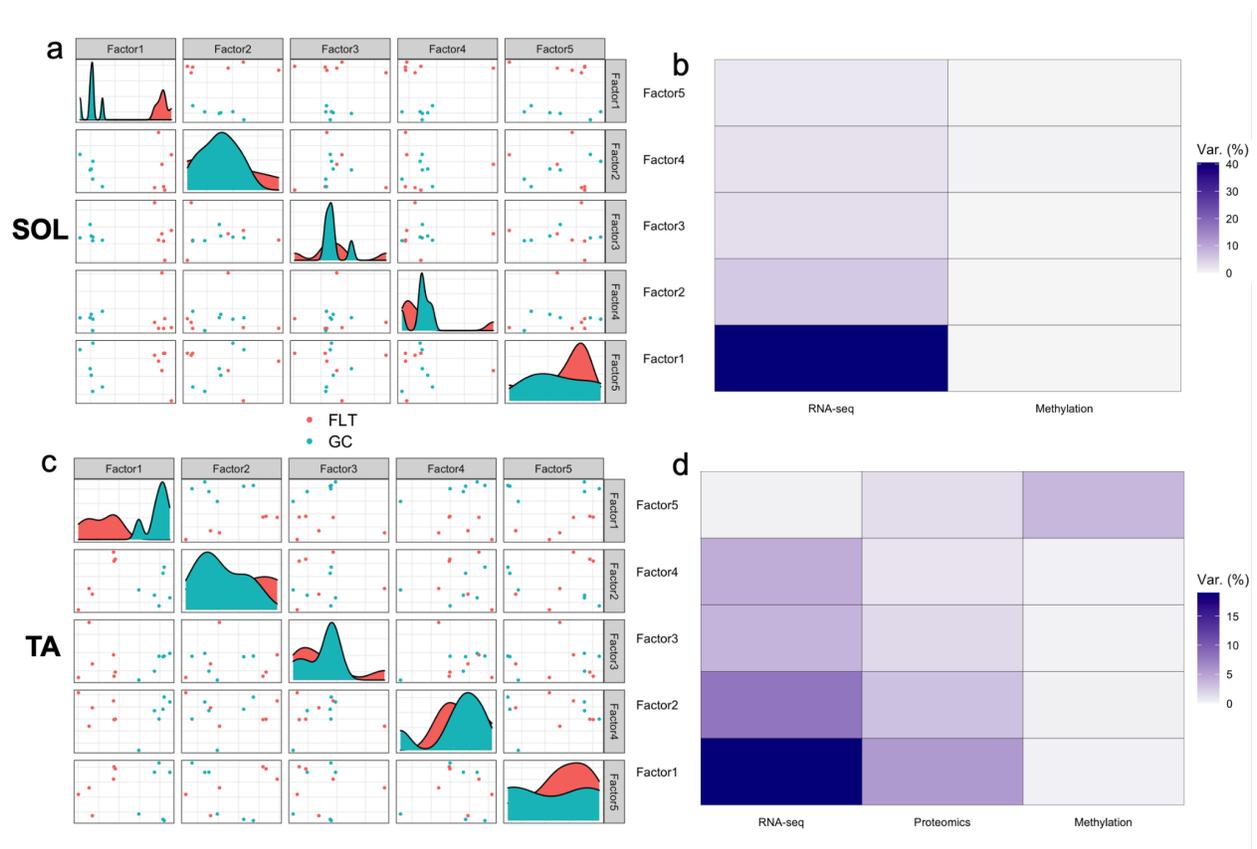

**Supplementary Figure 5. MOFA factors contributions. a)** Pair plots of the top 5 factors (by variance contribution) found by MOFA when using all available types of omics data for SOL. **b)** Contribution of each MOFA factor in explaining the % variance explained within each ome for SOL (OSD-104). **c)** Pair plots of the top 5 factors (by variance contribution) found by MOFA when using all available omes for TA. **d)** Contribution of each MOFA factor in explaining the percent variance explained within each ome for TA (OSD-105).

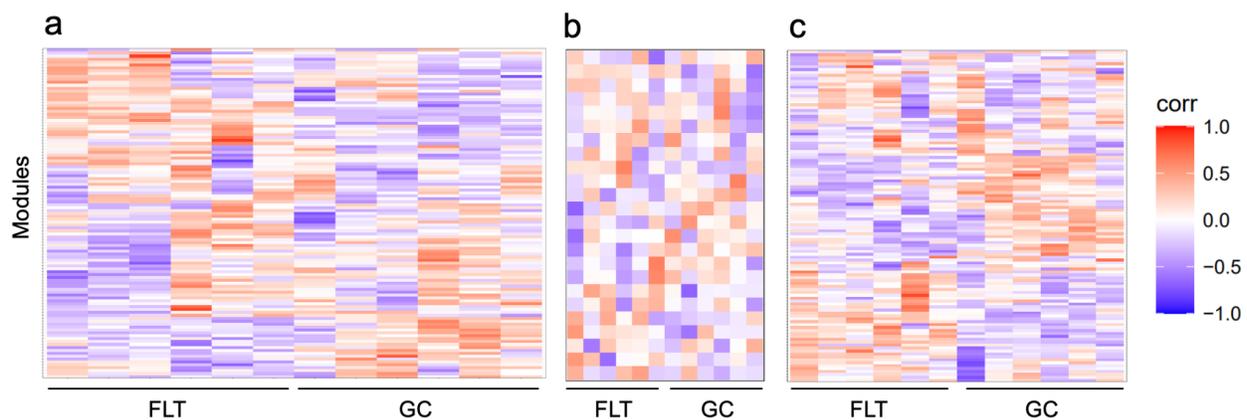

**Supplementary Figure 6. WGCNA clusters. A)** WGCNA clusters from TA RNA-seq data. **B)** WGCNA clusters from TA proteomics SOL data. **C)** WGCNA clusters from TA methylation data.

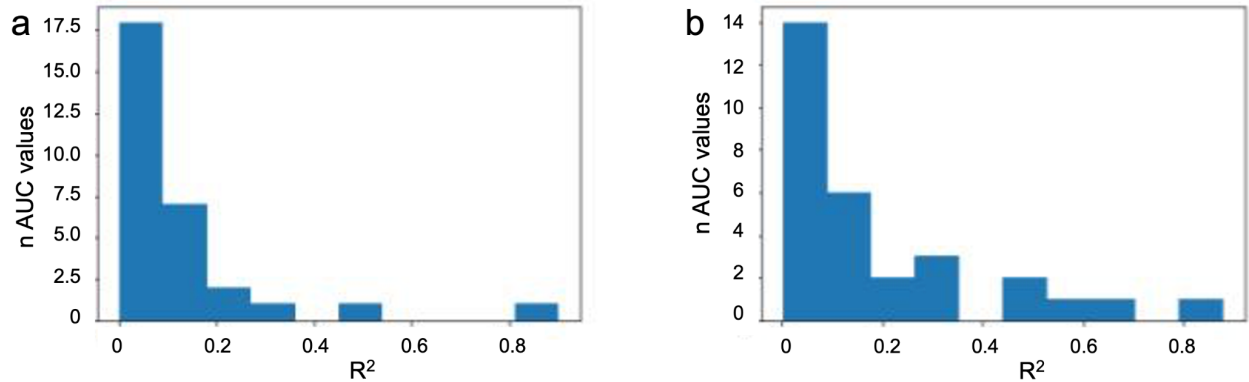

**Supplementary Figure 7. Calcium labeling assignment distributions.** The distribution of 30 random assignments of calcium AUC labels to **a)** OSD-104 SOL and **b)** OSD-105 TA mice. The x-axis is the resulting cross-validated $R^2$ scores when training QLattice with the labeling assignment.

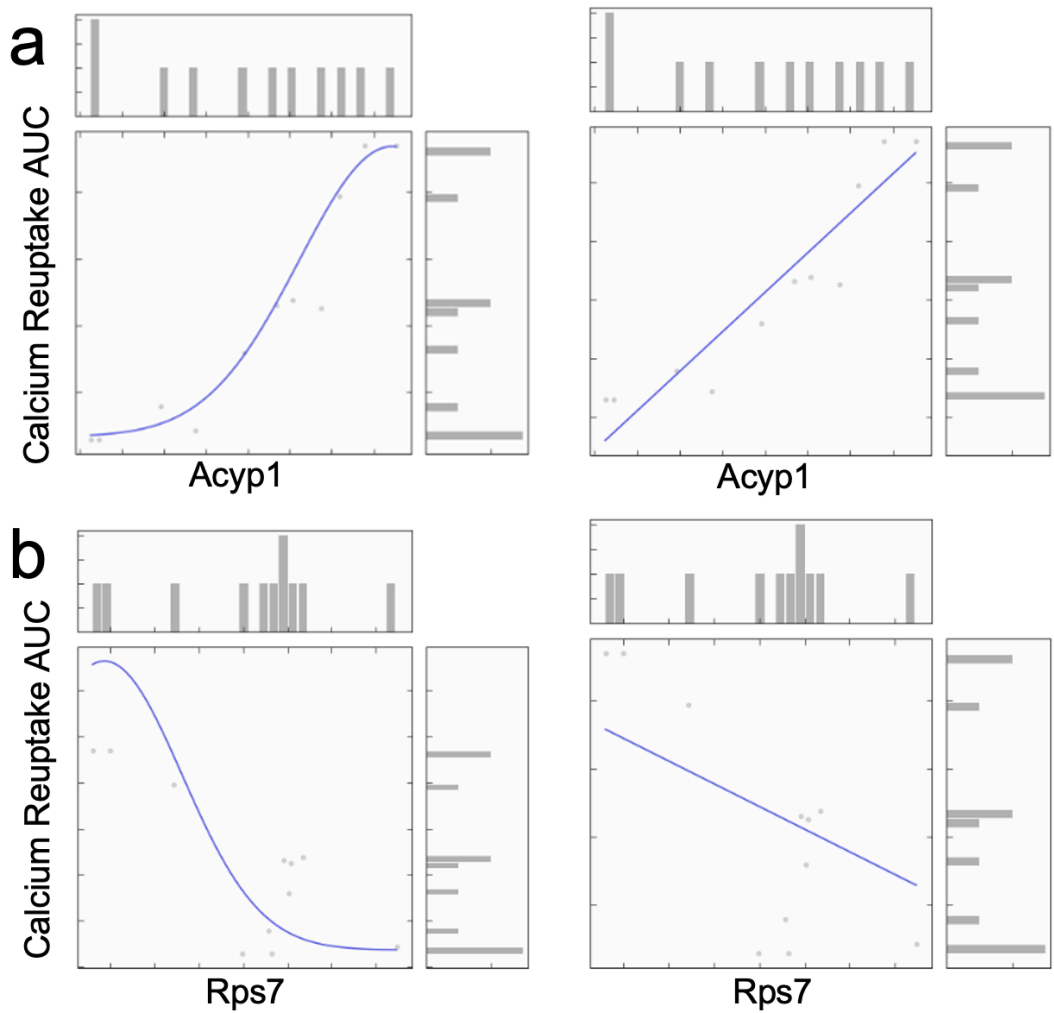

**Supplementary Figure 8. Relationships between protein expression and calcium reuptake AUC in QLattice models. a)** Representative examples of relationships between Acyp1 protein

and calcium reuptake AUC in QLattice models. **b)** Examples of relationships between Rps7 protein and calcium reuptake AUC.